\documentclass{webofc}
\usepackage[varg]{txfonts}   
\usepackage{hyperref}
\usepackage{subcaption}
\begin{document}
\title{Acceleration beyond lowest order event generation}
%
%
\subtitle{An outlook on further parallelism within \textsc{MadGraph5\_aMC@NLO}}

\author{\firstname{Zenny} \lastname{Wettersten}\inst{1,2}\thanks{\email{zenny.wettersten@cern.ch}}\and
        \firstname{Olivier} \lastname{Mattelaer}\inst{3} \and
        \firstname{Stefan} \lastname{Roiser}\inst{1} \and
        \firstname{Robert} \lastname{Schöfbeck}\inst{2} \and
        \firstname{Andrea} \lastname{Valassi}\inst{1} 
}

\institute{
           CERN
\and
           HEPHY
\and
        UCLouvain 
          }

\abstract{%
  An important area of high energy physics studies at the Large Hadron Collider (LHC) currently concerns the need for more extensive and precise comparison data. Important tools in this realm are event reweighing and evaluation of more precise next-to-leading order (NLO) processes via Monte Carlo event generators, especially in the context of the upcoming High Luminosity LHC. Current event generators need to improve throughputs for these studies. \textsc{MadGraph5\_aMC@NLO (MG5aMC)} is an event generator being used by LHC experiments which has been accelerated considerably with a port to GPU and vector CPU architectures, but as of yet only for leading order processes. In this contribution a prototype for event reweighing using the accelerated \textsc{MG5aMC} software, as well as plans for an NLO implementation, are presented.
}
\maketitle
\section{Introduction}
\label{sec:introduction}
As Moore's law's death throes echo throughout the world of high performance computing, a need for optimisation in hardware and software alike grows. Within high energy physics (HEP) this becomes especially apparent as we approach the era of the High Luminosity Large Hadron Collider (HL-LHC), where experimental precision and consequently both experimental and simulated measurements are expected to increase by an order of magnitude \cite{Bruning:2015dfu,Rossi:2019swj,Software:2815292,Collaboration:2802918}. Many different treatments for these difficulties are being studied, largely organised under the umbrella of the HEP Software Foundation \cite{Albrecht_2019,stewart2022hep}, but here we consider exclusively event generation, and particularly \textit{event reweighting}.

Event generation is the first step in HEP simulation, where process cross sections are evaluated and relevant instances of that process are stochasticly generated to be used for later stages of simulation. In total, event generation currently makes up $\sim5-15$\% of CPU hours at experiments \cite{HEPSoftwareFoundation:2020daq,HSFPhysicsEventGeneratorWG:2020gxw}. Due to the embarrassingly parallel and non-divergent nature of event generation, it is a prime candidate for exploring parallel architectures such as vectorised CPUs and GPUs, and over the last few years we have been working on porting leading order (LO) event generation within the \textsc{MadGraph5\_aMC@NLO} (\textsc{MG5aMC}) \cite{Alwall:2014hca} framework to such systems. Here, we consider what developments to pursue in the future as the accelerated \textsc{MG5aMC} port \textsc{(aMGaMC)} \cite{Valassi:2021ljk,Valassi:2022dkc,Valassi:2023yud} approaches an alpha release.

\section{Leading order event reweighing}
\label{sec:loreweighting}

Particle collision simulations factorise \cite{Mattelaer:2016gcx} into stages, and it is often unnecessary to resimulate samples for distinct physics models. Instead, it suffices to regenerate events under the new physics model, and simulations can be recycled to accommodate these new physics. Such reuse allows for more extensive work in fitting physics model parameters to experimental observations. Furthermore, scattering amplitudes factorise from MC event weights. Rather than regenerating events, amplitudes can be reevaluated under the assumption of the new model, and event weights \textit{reweighted} by this factor.

\subsection{Event generation and weights}
\label{sec:weights}

Given the inherently stochastic nature of particle physics, Monte Carlo (MC) methods are a natural fit for event generation. Normalising weights such that the total cross section is the sum of all event weights\footnote{As long as the normalisation is kept in mind, this choice is arbitrary.}, the weights $W$ of LO events are given by \cite{Mattelaer:2016gcx}
\begin{align}
\label{eq:weight}
    W = \left| M \right|^2 \; \left( \prod_i f_i (x_i, k^2 ) \right) \; \Omega_{PS},
\end{align}
with $|M|^2$ the scattering amplitude\footnote{Commonly referred to as \textit{matrix element}, \textit{matrix element squared}, \textit{amplitude}, or \textit{differential cross section}.} of the event, $f_i (x_i, k^2)$ the parton distribution function of parton $i$ with momentum fraction $x_i$ at renormalisation scale $k^2$, and $\Omega_{PS}$ the phase space measure. Defined as such, the total cross section $\sigma$ is
\begin{align}
    \sigma = \int \left| M \right|^2 \; \left( \prod_i f_i (x_i, k^2 ) \right) \; d \Omega \approxeq \sum_k W_k,
\end{align}
with $d \Omega$ the full angle differential, or equivalently the phase space measure. However, as mentioned, HEP events are stochastic; it is impossible to directly observe a total cross section, making the use of MC methods obvious: As only singular events can be observed in experiments, measurements naturally come about in a manner equivalent to a MC phase space distribution. For simulation purposes, MC methods are equivalent to real-world observations.

\subsection{Event reweighting}

Looking at \eqref{eq:weight}, the scattering amplitude is the sole factor dependant on internal physics of the interaction\footnote{The parton distribution functions depend only on the external partons, and the phase space measure is a volume.}. Consequently, if the physics model is modified,
\begin{align}
\label{eq:rwgt}
    |M|^2 \rightarrow |M'|^2 \; \implies \;  W \rightarrow W' = \frac{|M'|^2}{|M|^2} W,
\end{align}
i.e. so long as the relevant phase space for the new physics is a subset of the original one (ensuring $|M|^2 \neq 0$), the new event weight can be determined by refactoring the original weight with the new amplitude. This process is known as event reweighting, and although we treat reweighting an LO process to a modified LO process there are other use cases, e.g.\ reweighting LO events to NLO to increase precision without full NLO event generation \cite{HSFPhysicsEventGeneratorWG:2020gxw}.

\section{Parallel event generation with \textsc{MG5aMC}}
\label{sec:parallelgen}
\begin{figure}[t]
\centering
\sidecaption
\includegraphics[width=9.5cm,clip]{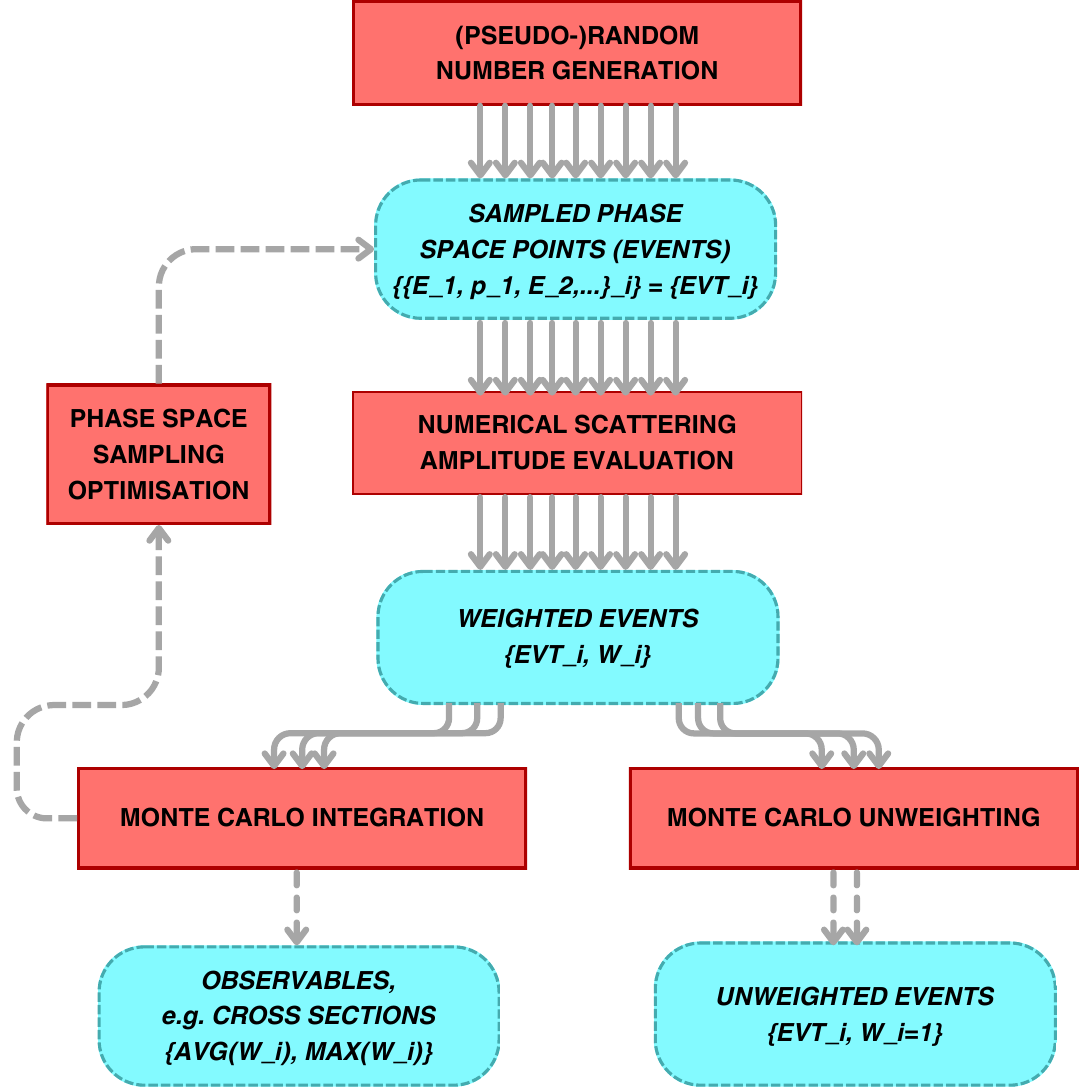}
\caption{Flowchart detailing the general structure of a scattering element generator. Red rectangles with right-angled corners correspond to numerical operations, while turquoise rounded nodes indicate their returned structures. Arrow abundance between nodes roughly hints at the amount of data in arguments and return values, save for all weighted events returned from the scattering amplitude evaluation being used for integration or unweighting. Returned observables and unweighted events can then be used by further simulation software.}
\label{fig:evgen}
\end{figure}
Due to the embarrassingly parallel and non-divergent nature of event generation, it makes a good candidate for heterogeneous computing in HEP. Over the last few years there has been work in porting \textsc{MG5aMC} event generation to the CUDA and SYCL standards \cite{Valassi:2021ljk,Valassi:2022dkc,Valassi:2023yud}, currently nearing release for LO event generation. Although we here primarily consider further development, a quick overview of the current framework is given below to contextualise work on parallel event reweighting and parallel NLO event generation.

\subsection{LO parallelisation}
\label{sec:loparallel}
The \textsc{MG5aMC} framework evaluates scattering amplitudes in terms of \textit{helicity amplitudes}, using ALOHA-generated HELAS-based code \cite{report:helas,art:aloha}. Event-level parallelism is an ``appropriate approach'' for acceleration \citep{HSFPhysicsEventGeneratorWG:2020gxw}, and exactly what has been implemented: The code structure is inherited from the \textsc{MadEvent} event generator, but amplitudes\footnote{Note that \textit{only} scattering amplitude evaluations are parallelised here. Other event-specific parts, such as random number generation, are kept on the host. Further points of acceleration are being investigated, but are not a priority.} are evaluated with vectorised C++ code on CPUs, and using the CUDA API on GPUs.

An overview of the MC event generator structure is provided in Fig.\ \ref{fig:evgen}, where the green bubble specifies the part event-level parallelism applies to. Note that \textsc{MG5aMC} is a meta-program, a \textit{code generator}, rather than a singular implementation. \textsc{MG5aMC} generates, exports, and runs a program for considered physics processes. Particularly, scattering amplitude evaluations are called as external subroutines distinct from phase space integration. Consequently, it is simple (although not easy) to replace these with e.g.\ vectorised routines. Scattering amplitudes being the primary bottleneck in event generation, this is the consideration for acceleration in \textsc{aMGaMC}.

\subsection{Parallel LO scattering amplitudes and reweighting}
As detailed in Sec.\ \ref{sec:loparallel}, not only are scattering amplitude routines distinct from other event generation steps, but \textsc{MG5aMC} and \textsc{aMGaMC} generate them independently. In light of Sec.\ \ref{sec:loreweighting}, where event reweighting was shown to only use amplitude evaluation, these routines can be repurposed for reweighting. This is considered here: The usage of HELAS-like CUDA code, in line with the native \textsc{MG5aMC} reweighting module \cite{Artoisenet:2010cn}.

Although \textsc{aMGaMC} does not yet support out-of-the-box event generation --- generated code takes a posteriori modifications --- standalone processes (i.e. amplitude generation, the green bubble in Fig.\ \ref{fig:evgen}) \textit{do} work. Using this standalone output, a minimal program calling an in-house generic reweighting library\footnote{This library is expected to see an official release before the end of the year. We forego further details here.} can perform event reweighting on a provided event set.

This comes with a caveat: There exists no interface between these parts yet. With \textsc{aMGaMC}, the necessary amplitude subroutines can largely\footnote{\textsc{aMGaMC} does not yet support fully generic beyond-the-Standard Model process generation.} be created; our reweighting library reads and runs events through generated subroutines; and a separate program calls these libraries to perform the reweighting.

Once set up and compiled, reweighting goes roughly as follows:
    \textbf{1)} Input reweight parameters;
    \textbf{2)} Parse events;
    \textbf{3)} Evaluate original scattering amplitudes;
    \textbf{4)} Overwrite physics parameters with new values;
    \textbf{5)} Evaluate new amplitudes;
    \textbf{6)} Evaluate new weight with eq. \eqref{eq:rwgt};
    \textbf{7)} Repeat steps $4 - 6$ for all parameter sets, and;
    \textbf{8)} Return new weights.
We note that this procedure only holds for parameters that can be modified at runtime, making reweighting between structurally different amplitude routines tedious (albeit possible).

\subsection{Results}
\label{sec:results}
\begin{table}[t]
    \centering
    \begin{tabular}{|c|c|}\hline
        \textbf{Process} & $e^+ e^- \rightarrow 5 \gamma$ (SM) \\ \hline
         \textbf{CPU} & Intel Core i5-1145G7 \\ \hline
         \textbf{GPU} & NVIDIA A100 PCIe 40GB \\ \hline
    \end{tabular}
    \caption{Details for the time measurements displayed in Figures \ref{fig:scaleNEvts} and \ref{fig:scaleNPars}, showing the process considered and hardware used for the two implementations. Note that all reweighting is done on within the standard model, setting the electroweak coupling constant to be a randomly determined non-zero value for each reweight iteration.}
    \label{tab:details}
\end{table}
\begin{figure}[t]
    \centering
    \begin{subfigure}[b]{0.49\textwidth}
    \includegraphics[width=\textwidth,clip]{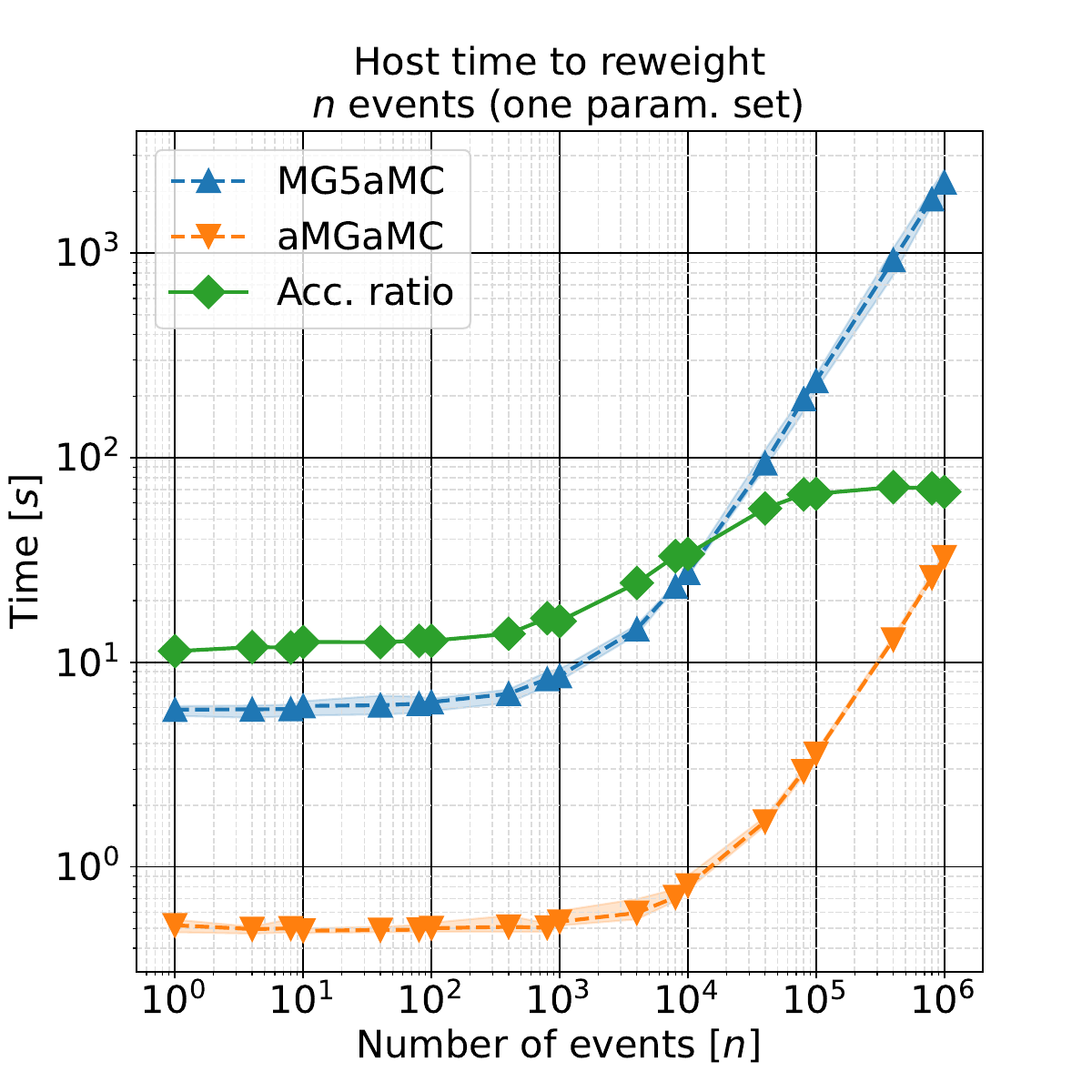}
    \end{subfigure}
    \hfill
    \begin{subfigure}[b]{0.49\textwidth}
    \includegraphics[width=\textwidth,clip]{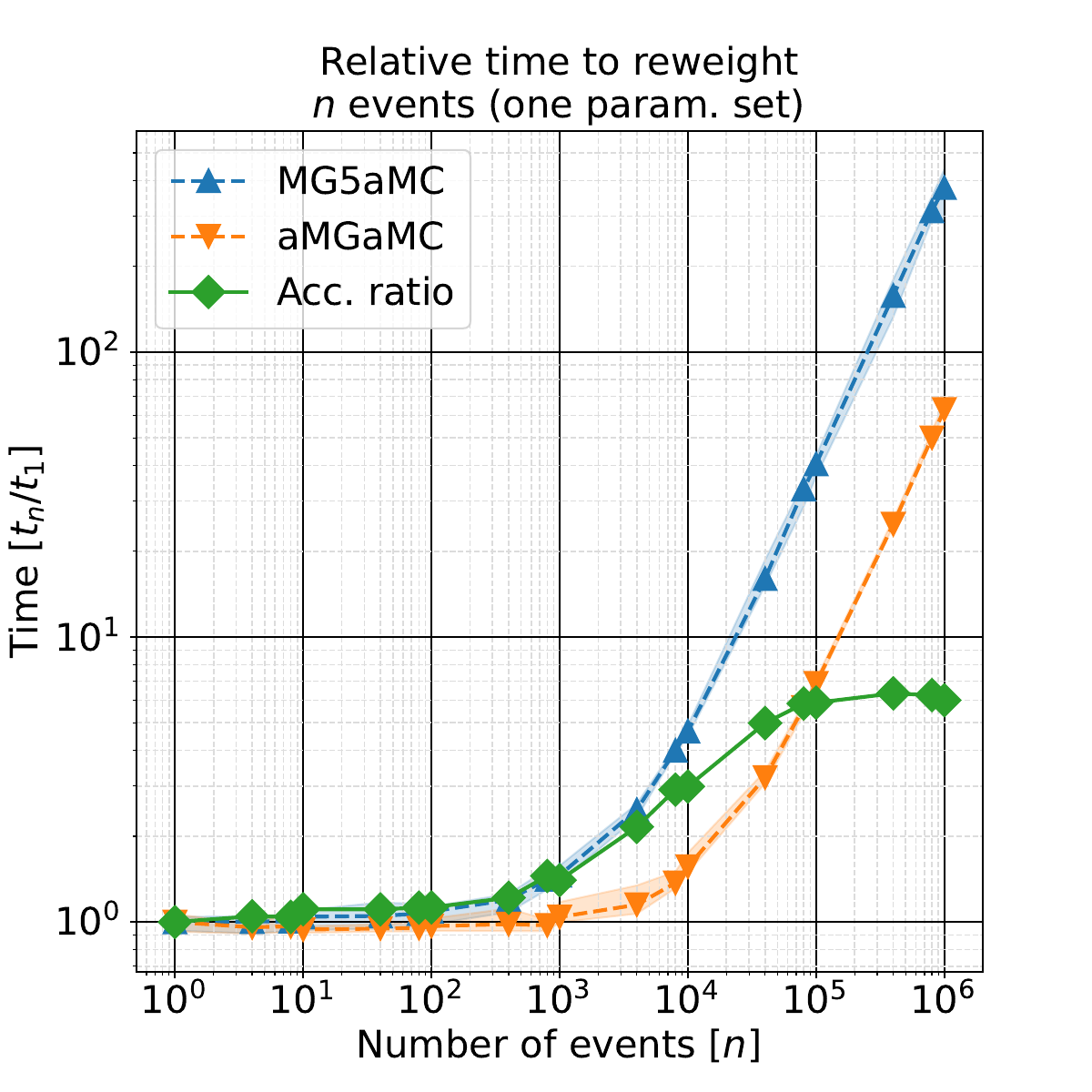}
    \end{subfigure}
    \caption{Comparison of real (left) and relative (right) host (CPU) time to reweight $n$ events for one set of parameters for \textsc{MG5aMC} and \textsc{aMGaMC}. Average time over five measurements is shown, with min- and maximum value for each $n$ denoted by coloured regions. In the right plot, times have been independently normalised to the time $t_1$ it takes to reweight one event once. The green solids show the ratio between the two dashed lines. See Table \ref{tab:details} for details on the reweight procedure.}
    \label{fig:scaleNEvts}
\end{figure}
\begin{figure}[t]
    \centering
    \begin{subfigure}[b]{0.49\textwidth}
    \includegraphics[width=\textwidth,clip]{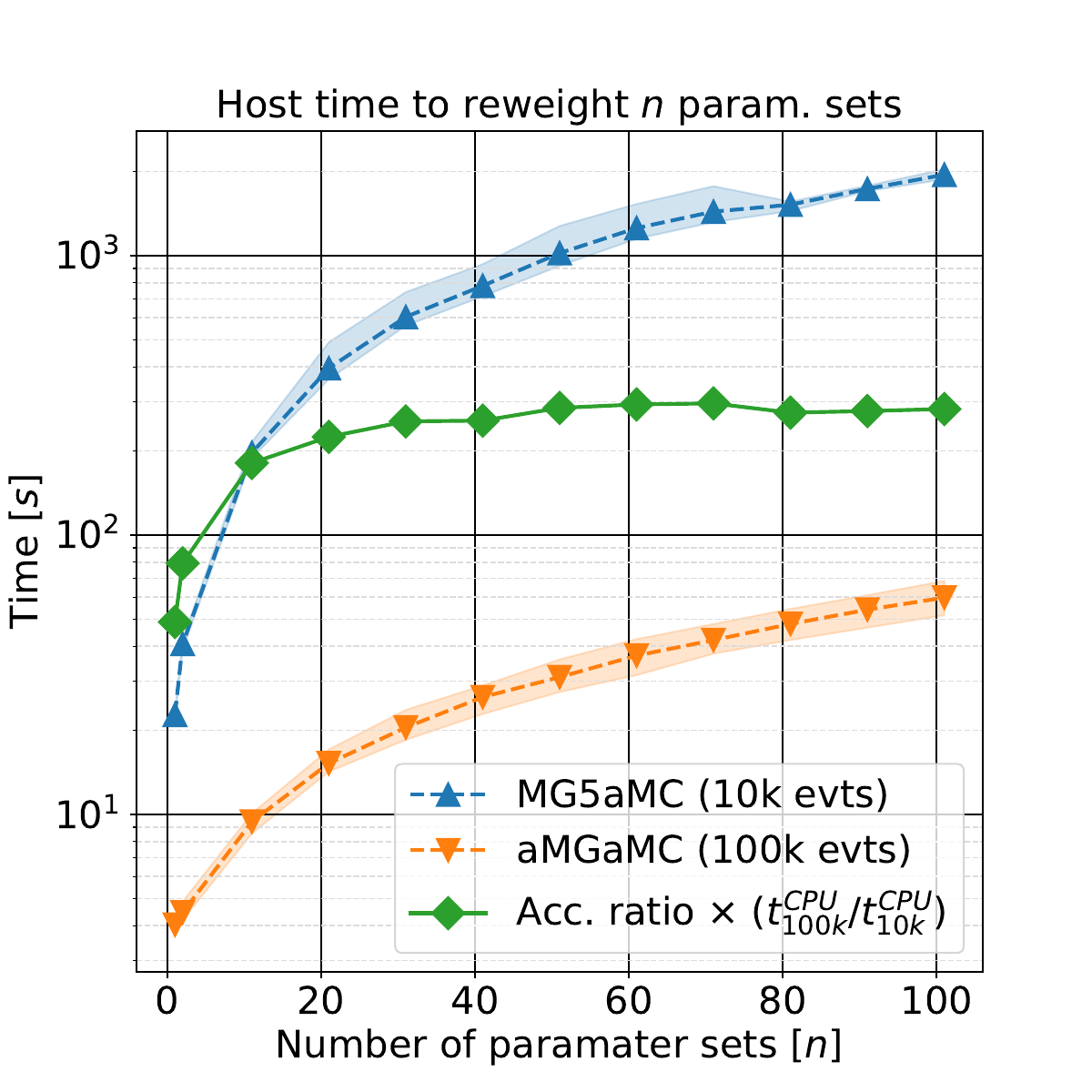}
    \end{subfigure}
    \hfill
    \begin{subfigure}[b]{0.49\textwidth}
    \includegraphics[width=\textwidth,clip]{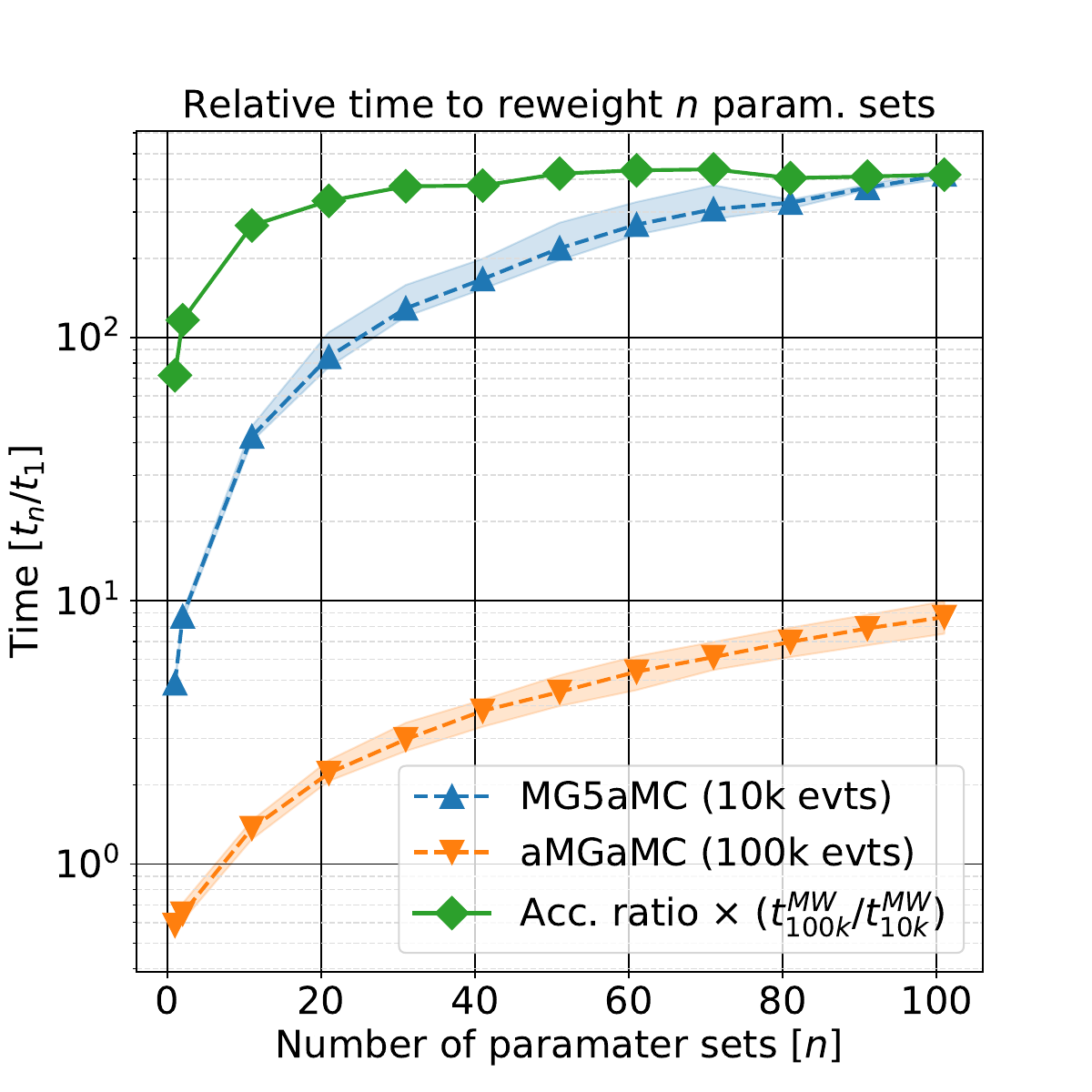}
    \end{subfigure}
    \caption{Comparison of real (left) and relative (right) host (CPU) time to reweight $10^4$ (\textsc{MG5aMC}) or $10^5$ (\textsc{aMGaMC}) events for $n$ parameter sets. Average time over five measurements is shown, with min- and maximum value for each $n$ denoted by coloured regions. In the right plot, the normalisation shown in Fig.\ \ref{fig:scaleNEvts} (right plot) is used. The green solid lines show the ratio between the dashed lines, multiplied by a factor $\sim8.66$ longer it takes \textsc{MG5aMC} to reweight $10^5$ events than $10^4$. See Table \ref{tab:details} for details on the reweight procedure.}
    \label{fig:scaleNPars}
\end{figure}
To evaluate the reweight acceleration of \textsc{aMGaMC}, two independent degrees of complexity need be considered: For statistics, it is necessary to increase the amount of events; for physics studies, it becomes important to consider more parameter sets. Practically, these are close to identical, as reweighting over multiple parameter sets means performing more amplitude evaluations for the same event. Runtimes are thus expected to rise linearly with both, at least beyond some minimum number of events (per parameter set) such that amplitudes dominate runtime. In Table \ref{tab:details}, the measurement details are shown\footnote{Note that the CPU used is consumer grade, making one-to-one runtime comparisons less relevant.}, and time measurements are plotted against number of events and of parameter sets in Figs.\ \ref{fig:scaleNEvts} and \ref{fig:scaleNPars}, respectively.

As can be seen in Figs.\ \ref{fig:scaleNEvts} and \ref{fig:scaleNPars}, runtimes increase linearly\footnote{Note that Fig.\ \ref{fig:scaleNEvts} has log-log graphs, whereas Fig.\ \ref{fig:scaleNPars} has log-linear graphs.} with reweight process complexity for both \textsc{MG5aMC} and \textsc{aMGaMC}. The left-hand graphs show total CPU times, while for the right-hand graphs the runtimes have been independently normalised to the time to reweight a single event for one parameter set. For small event sets, both implementation runtimes are dominated by (initialisation, I/O, etc.), making this choice unfit for \textsc{aMGaMC} due to the overhead occurring on the host rather than on the device, resulting in a bias against the GPU.

In Fig.\ \ref{fig:scaleNEvts}, similar runtime complexities are observed between the two implementations; initial domination by statistical effects, but a trend towards linear growth in the large-$n$ limit. The main difference is an initial constant runtime for \textsc{aMGaMC}, explained by two factors: Predominant overhead in the small-$n$ limit is transfer rate between host and device, which for small $n$ is roughly constant; and amplitude evaluation runtime will not increase until the device is saturated, which will happen at $\sim 10^4$ events. Acceleration from the GPU is depicted here by green lines, which show the ratio between the two implementation runtimes.
Fig.\ \ref{fig:scaleNPars}, which details the runtimes as functions of the number of parameter sets, shows a clear linear growth for both implementations. As mentioned, reweighting over several parameter sets is equivalent to reweighting more events when amplitude evaluation is the dominant runtime contribution, and for Fig.\ \ref{fig:scaleNPars} sufficiently many events per iteration have been taken for both implementations to land in the linear large-$n$ limit shown in Fig.\ \ref{fig:scaleNEvts}. Note that the green lines depicting the runtime ratio in Fig.\ \ref{fig:scaleNPars} have been multiplied by the runtime quotient $\sim 8.66$ between $10^5$ and $10^4$ events for \textsc{MG5aMC}, to account for these measurements being made for $10^4$ events rather than the $10^5$ events of \textsc{aMGaMC}. This is a logistical limitation in the runtime for such large event sets, but the factor used is biased towards \textsc{MG5aMC}\footnote{Most overhead (particularly I/O) depends only on the number of events, not on the number of parameter sets: Each event needs to be parsed only once, overcounting this overhead for each parameter set past the first one.}, and this factor is nevertheless not much smaller than the naive maximum factor $10$.
\section{Expectations for NLO parallelism}
While we approach an official LO \textsc{aMGaMC} release, we turn to the next major goal: NLO event generation. Work is just commencing to this end, but considerations of \textsc{aMGaMC} alongside the structure of NLO event generation in native \textsc{MG5aMC} can be presented.

In \textsc{MG5aMC}, NLO event generation uses the FKS subtraction scheme \citep{Frixione:1995ms,Frixione:1997np,Frederix:2018nkq}. We forego theoretical details here, but note that this splits NLO amplitudes into three distinct subsets:
\begin{description}
    \item[LO amplitudes.] In perturbation theory, higher order contributions are \textit{corrections} to LO amplitudes. Thus, LO amplitudes need to be evaluated also for NLO calculations.
    \item[Real emission amplitudes.] Here, additional unobserved particles are added to the interaction. These are \textit{tree-level}, and can be evaluated using the same machinery as LO amplitudes.
    \item[One-loop amplitudes.] These loop amplitudes in NLO corrections require integration over undetermined momenta for singular amplitudes. While many clever techniques for such evaluations exist, it requires different machinery to that at tree-level.
\end{description}

In porting LO event generation to parallel architectures the structures used for tree-level amplitudes have already been developed; we already have the tools to evaluate real emissions. The first two points for NLO event generation can be treated as LO multiprocesses, and although their NLO use has not been implemented, we expect few problems in doing so.

Loops, though, present an issue. Native \textsc{MG5aMC} calls external libraries to evaluate loop integrals \citep{Alwall:2014hca}; development of such GPU-compatible libraries is largely unaddressed as of yet\footnote{GPU-focused loop developments \citep{Yuasa_2013,Li:2015foa,SMIRNOV2016189,Winterhalder:2021ngy} largely focus on speeding up the loop integrals themselves, whereas we consider event-level parallelism.}. Additionally, depending on the numerical stability, it is at times necessary to evaluate loop integrals at quadruple precision, which no GPU or TPU manufacturer currently supports. Nevertheless, it appears that loops could be evaluated on GPUs for at least a large fraction of events.

\section{Outlook}
\label{sec:outlook}
Over the course of this paper, we have detailed expected difficulties in development of \textsc{aMGaMC} past LO event generation, as well as presented a simple comparison in how already supported development can be extended for purposes other than pure scattering amplitude generation, here using \textsc{aMGaMC} to parallelise scattering amplitudes for LO event reweighting. Although resulting runtimes are difficult to compare one-to-one, parallelism evidently makes the encroaching need for larger event sets in HEP less problematic.

Nevertheless, LO event generation parallelism will be insufficient to tackle computational needs in the coming decade. The difficulties in porting also NLO scattering amplitude generation are being investigated, and while the computation of one-loop amplitudes is predicted to pose a problem, we are optimistic. The architecture for parallel evaluation of tree-level diagram amplitudes already exists, and a first partial port of NLO event generation to \textsc{aMGaMC} could be done by performing these tree-level evaluations on vectorised architecture, keeping loop evaluations on the host.

Somewhat orthogonal to this development, we are working on running also LO event reweighting on heterogeneous architectures. A C++ library for parsing the LHE file format and interfacing with generic structures for performing event reweighting has been developed, and we hope to have it released publicly by the end of the year. Interfacing this library with the current pre-alpha release of \textsc{aMGaMC}, a test bed for LO event reweighting on GPUs has been established, demonstrating the sizeable acceleration attained with vectorisation and GPU parallelism. In Figs.\ \ref{fig:scaleNEvts} and \ref{fig:scaleNPars}, we display that in the high complexity limit, the \textsc{aMGaMC} reweighting implementation will be a sizeable factor faster than \textsc{MG5aMC}. The specific factor is, of course, hardware-dependent.

Although some complications must be dealt with prior to an official release of the \textsc{aMGaMC} plugin, deliberation on future development plans is becoming increasingly important. A favoured candidate is immediately moving onto NLO event generation, and here we have discussed some of the foreseen obstacles. However, other paths to success have revealed themselves as well: Here we have treated event reweighting, which allows for recycling generated event sets to probe physics parameters without the need to regenerate or resimulate events. Continued effort in exploring attainable benefits in heterogeneous architectures and in how to apply these efforts in multiple synergistic directions appears to be not only effective, but necessary, for the future of particle physics.

\bibliography{bibliography}

\end{document}